\documentclass[prl,showpacs,floatfix,twocolumn]{revtex4}

\usepackage{graphicx}
\usepackage{amsmath}

\begin{document}

\title{Quantum vortex tunneling: Microscopic theory and application to $d$-wave superconductors}
\author{Ari Mizel}
\affiliation{Department of Physics and Materials Research Institute, Penn State University, University Park, Pennsylvania 16802, U.S.A.}
\date{\today} 

\begin{abstract}
We present a microscopic approach to the quantum tunneling of vortices.
The formalism characterizes the rate at which a many-body
superconducting state with a vortex in one location makes a transition
to a second many-body superconducting state with a vortex in a second
location.  The theory is utilized to study the effect of a $d$-wave
order parameter upon the directionality of vortex tunneling.
\end{abstract}

\maketitle

For decades, researchers have studied the motion of vortices in
type-II superconductors \cite{Kim}.  This motion determines
fundamental characteristics of superconductors, including critical
current and mixed-state resistivity \cite{Tinkham}, which often
dictate their technological usefulness.  Because of the complicated
nature of a vortex in a superconductor, however, many aspects of the
theoretical understanding of vortex motion have remained incomplete.
In particular, the theory of quantum tunneling of vortices
\cite{Tunneling,Simanek} has been phenomenological in character,
treating a vortex as a point-particle governed by an effective action.
In this paper, we approach the problem of vortex quantum tunneling
from a fully microscopic quantum mechanical perspective.  Our
calculations are based on the Nambu-Gorkov equations of
superconductivity; we do not require effective parameters and do not
assume that the vortex can be treated as a point-particle.  Because of
its microscopic foundation, the theory (i) is capable of addressing
qualitative phenomena that are beyond the reach of effective models
and (ii) provides a foundation for quantitative, first principles
analysis of vortex motion in real materials.  We focus in this paper
upon non-dissipative motion, a topic that has elicited sustained
theoretical concern even without the additional complication of
dissipation\cite{Nondiss,Niu} (dissipation would be incorporated into
the theory in the spirit of \cite{Caldeira}).  As a first application,
we study the effects of the $d$-wave nature of the order parameter in
high-temperature superconductors \cite{Dwave}.  Our calculations
predict that $d$-wave symmetry suppresses tunneling along the nodes of
the order parameter.  This effect cannot be captured even in principle
within the existing point-particle theories \cite{anisotropy}.

We begin our formulation by introducing the state $\left| v_N \right>$
that describes a superconductor with $N$ electrons forming a vortex at
the origin.  For simplicity, we assume that the electrons are confined
to 2 dimensions, which is appropriate for a superconducting thin film
or possibly for a layered superconductor.  The extension to 3
dimensions is straightforward.  The microscopic Hamiltonian of the
system is $H=H_o+V$ where $H_o$ describes electrons in an ionic
potential and a homogeneous magnetic field, undergoing an effective
electron-electron interaction that gives rise to superconductivity.  A
separate pinning potential $V$ results from microscopic impurities or
dislocations.

We consider the dynamics of this Hamiltonian in a basis of states of
the form $\left| v_N(\mbox{\boldmath$\rho$}) \right> \equiv
T(\mbox{\boldmath$\rho$}) \left| v_N \right>$, where the operator
$T(\mbox{\boldmath$\rho$})$ translates the vortex from the origin to
the position $\mbox{\boldmath$\rho$}$.  It is important to emphasize
that we are {\em not} making a phenomenological point-particle
\cite{Tunneling} assumption about the vortex coordinate
$\mbox{\boldmath$\rho$}$ when we deal with states of the form $\left|
v_N(\mbox{\boldmath$\rho$})\right>$; the vortex states are the lowest
energy states of a superconductor in a magnetic field \cite{deGennes}
and simply compose the low-energy Hilbert space of the microscopic
Hamiltonian.  Physically, this basis is a natural choice for
describing a superconductor with a moving vortex.  To compute the
behavior of a time-dependent vortex state \mbox{$\left| \Psi (t)
\right> = \sum _{\mbox{\boldmath$\rho$}^{\prime}}
\Psi(\mbox{\boldmath$\rho$}^{\prime},t) \left|
v_N(\mbox{\boldmath$\rho$}^{\prime}) \right>$}, in the usual way we
first compute the eigenstates \mbox{$\left| \psi \right> = \sum
_{\mbox{\boldmath$\rho$}^{\prime}}
\psi(\mbox{\boldmath$\rho$}^{\prime}) \left|
v_N(\mbox{\boldmath$\rho$}^{\prime}) \right>$} of the time-independent
Schr\"odinger equation \mbox{$H_o + V - E \left| \psi \right> = 0$}.
In our basis, this equation is
\begin{equation}
\label{schroeqn}
\sum _{\mbox{\boldmath$\rho$}^{\prime}} \left< v_N(\mbox{\boldmath$\rho$}) \right| H_o + V - E \left| v_N(\mbox{\boldmath$\rho$}^{\prime}) \right> \psi(\mbox{\boldmath$\rho$}^{\prime}) = 0 \; .
\end{equation}
Before we can contemplate solving this equation, the essential
challenge is to evaluate the matrix elements $\left<
v_N(\mbox{\boldmath$\rho$}) \right| H_o + V \left|
v_N(\mbox{\boldmath$\rho$}^{\prime}) \right>$ and the overlap $\left<
v_N(\mbox{\boldmath$\rho$}) \right. \left|
v_N(\mbox{\boldmath$\rho$}^{\prime}) \right>$, keeping in mind that
the states in our basis are not orthogonal.  These matrix elements
seem to require $2N$ dimensional many-body integrals, which can become
 utterly intractable for even relatively small $N$.  In the
following, we provide tractable forms for them by describing $\left|
v_N(\mbox{\boldmath$\rho$}) \right>$ using the mean-field,
Nambu-Gorkov equations of superconductivity \cite{Gorkov}.  Of course,
a solution to (\ref{schroeqn}) is a superposition \mbox{$\left| \psi
\right> = \sum _{\mbox{\boldmath$\rho$}^{\prime}}
\psi(\mbox{\boldmath$\rho$}^{\prime}) \left|
v_N(\mbox{\boldmath$\rho$}^{\prime}) \right>$} of mean-field states,
so that our analysis goes well beyond mean-field theory.

To employ Nambu-Gorkov theory, we introduce the state $\left| v
\right>$ of a superconductor that has a vortex at the origin and is in
contact with a particle bath.  This state can be well
characterized using mean-field theory and the $N$ electron state
$\left| v_N\right>$ can be obtained from it by projection $\left| v_N
\right> = P_N \left| v \right>$.  The mean-field description of
$\left| v \right>$ is contained in the Green's functions,
$G_{s,s^{\prime}}({\mathbf r},{\mathbf r}^{\prime},t) = - \left< v
\right| T \{ d _{{\mathbf r},s}(t),d^{\dagger} _{{\mathbf
r^{\prime}},s^{\prime}}(t^{\prime}=0)\} \left| v \right>$ which are
defined using a particle-hole transformation from the usual electron
destruction operators to $d_{{\mathbf r},s = 1} = c_{{\mathbf r},s =
1} $ and $d_{{\mathbf r},s = -1} = c^{\dagger}_{{\mathbf r},s = -1}$.
After Fourier transforming in time, one finds
\cite{Gorkov}
\begin{equation}
\label{greeneqn}
\left( \begin{array}{cc} K - E& \hat{\Delta}  \\ \hat{\Delta} ^* & -K^* - E\end{array} \right) {\cal G} = i \hbar \delta({\mathbf r}-{\mathbf r^{\prime}}) \left( \begin{array}{cc} 1  & 0\\ 0 & 1 \end{array} \right)
\end{equation}
where $K = \frac{1}{2m}({\mathbf p} + \frac{|e|}{c} {\mathbf A})^2 +
U({\mathbf r}) - E_F$ and ${\cal G}$ is a 2 by 2 matrix with elements
$G_{s,s^{\prime}} ({\mathbf r},{\mathbf r^{\prime}},E).$ The crystal
potential $U({\mathbf r})$ includes the ionic potential and
electron-electron Hartree interactions.  The superconducting
order parameter operator $\hat{\Delta}$ is defined by $\hat{\Delta} \;
G_{s,s^{\prime}}(\tilde{\mathbf r},{\mathbf r^{\prime}},E) = \int d
\tilde{\mathbf r} \; \Delta({\mathbf r},\tilde{\mathbf r}) \;
G_{s,s^{\prime}}(\tilde{\mathbf r},{\mathbf r^{\prime}},E)$ and should
be computed self-consistently.  This equation can be solved
numerically on a finite real space lattice, in which case the Green's
function can be regarded as a matrix.

The vortex state $\left| v_N(\mbox{\boldmath$\rho$}) \right>$ is
 obtained by translating $\left| v_N \right>$, which
contains a vortex at the origin.  Because of the presence of an
external, homogeneous magnetic field in the Hamiltonian, one effects the translations using magnetic translation
operators \cite{Zak,Morandi}.  Magnetic translation operators are
usually defined with respect to single-particle wavefunctions; in the
Landau gauge ${\mathbf A}({\mathbf r}) = - B y \hat{x}$, we have the
definition $\tau(\mbox{\boldmath$\rho$}) \psi({\mathbf r}) = e^{i
(\pi B/\Phi_o) \rho _y (x - \rho _x /2)}\psi({\mathbf
r}-\mbox{\boldmath$\rho$})$ where $\Phi_o = h c/2|e|$ is the
superconducting flux quantum \cite{Morandi}.  This operator must be
generalized in order to translate a many-body vortex state properly.
A suitable generalization is
\begin{eqnarray}
T(\mbox{\boldmath$\rho$}) &= & \Pi _{{\mathbf R},S} (C _{{\mathbf
R},S} C ^\dagger _{{\mathbf
R},S} + c^\dagger _{{\mathbf
R},S} C _{{\mathbf
R},S}) \notag \\
& \times & \Pi _{{\mathbf r},s} (c_{{\mathbf r}-\mbox{\boldmath$\rho$},s}c^\dagger_{{\mathbf r}-\mbox{\boldmath$\rho$},s} + e^{i (\pi B/\Phi_o) \rho _y (x -
\rho _x /2)} C^{\dagger}_{{\mathbf r},s}c_{{\mathbf r}-\mbox{\boldmath$\rho$},s}). \notag
\end{eqnarray}
In this expression we have introduced fermionic operators of the form
$C^\dagger$ that create electrons in an artificial, auxillary space.
Such operators are needed because ``holding'' space is required when
exchanging a collection of objects -- to exchange the balls in two
boxes, a third ``holding'' box is required temporarily.  In other
words, one should not shift an electron from ${\mathbf r}$ to
${\mathbf r}+\mbox{\boldmath$\rho$}$ without first placing an electron
already in ${\mathbf r}+\mbox{\boldmath$\rho$}$ into a holding box.
One can verify that $T(\mbox{\boldmath$\rho$}) c^{\dagger}_{{\mathbf r}_1 -\mbox{\boldmath$\rho$},s_1}c^{\dagger}_{{\mathbf r}_2-\mbox{\boldmath$\rho$},s_2}...c^{\dagger}_{{\mathbf r}_n-\mbox{\boldmath$\rho$},s_n} = e^{i (\pi B/\Phi_o) \rho _y (x_1 - \rho _x /2)} c^{\dagger}_{{\mathbf r}_1,s_1}...e^{i (\pi B/\Phi_o) \rho _y (x_n - \rho _x /2)}c^{\dagger}_{{\mathbf r}_n,s_n}$, as expected.  If needed, the number projection operator $P_N$ can be conveniently incorporated into
$T(\mbox{\boldmath$\rho$})$ as follows
\begin{eqnarray*}
\lefteqn{T(\mbox{\boldmath$\rho$})P_N = \int \frac{d \chi}{2\pi} e^{-i N \chi} \Pi _{{\mathbf R},S} (C _{{\mathbf R},S} C ^\dagger _{{\mathbf R},S} + c^\dagger _{{\mathbf R},S} C _{{\mathbf R},S})} \notag \\
& \times & \Pi _{{\mathbf r},s} (c_{{\mathbf r}-\mbox{\boldmath$\rho$},s}c^\dagger_{{\mathbf r}-\mbox{\boldmath$\rho$},s} + e^{i (\chi + (\pi B/\Phi_o) \rho _y (x - \rho _x /2))} C^{\dagger}_{{\mathbf r},s}c_{{\mathbf r}-\mbox{\boldmath$\rho$},s}).
\end{eqnarray*}

After applying the particle-hole transformation to this operator, one can use Wick's theorem \cite{condmat} to evaluate the overlap matrix element $\left< v_N (\mbox{\boldmath$\rho$}) \right. \left| v_N (\mbox{\boldmath$\rho$}^{\prime})\right> =  \left< v \right|T^\dagger(\mbox{\boldmath$\rho$})T(\mbox{\boldmath$\rho$}^\prime)P_N \left| v \right>$ that appears in equation (\ref{schroeqn}).  Up to an unimportant phase factor that can be absorbed into the vortex state, the result is
\begin{equation}
\left< v_N (\mbox{\boldmath$\rho$}) \right. \left| v_N (\mbox{\boldmath$\rho$}^\prime)\right>  =  \int \frac{d\chi}{2\pi}  e^{i(N_o/2 -N)\chi}   \hbox{det} \; {\cal M} (\chi). \label{overlap}
\end{equation}
where $N_o$ is the number of single-particle orbitals in the system and
\begin{widetext}
\begin{equation*}
{\cal M}(\chi) _{({\mathbf r},s);({\mathbf r^{\prime}},s^{\prime})} =  - G_{s,s^{\prime}}({\mathbf r}-\mbox{\boldmath$\rho^\prime$},{\mathbf r}^{\prime}-\mbox{\boldmath$\rho^\prime$},t=0^+) + G_{s,s^{\prime}}({\mathbf r}-\mbox{\boldmath$\rho^\prime$},{\mathbf r}^\prime - \mbox{\boldmath$\rho$},t=0^-)e^{is^{\prime}\left(\chi + (\pi B/\Phi_o) ({\rho _y}^\prime (x^\prime - {\rho _x}^\prime /2) - \rho _y (x^\prime - \rho _x/2))\right)}
\end{equation*}
\end{widetext}
is an $N_o \times N_o$ matrix indexed by row $({\mathbf r},s)$ and column $({\mathbf r^{\prime}},s^{\prime})$.

Note that equation (\ref{overlap}) involves only determinants and a single dimensional integral.  In place of a totally unmanagable $2N$ dimensional many-body integral for $\left< v_N (\mbox{\boldmath$\rho$}) \right. \left| v_N (\mbox{\boldmath$\rho$}^\prime)\right>$, it provides a computationally tractable expression that we evaluate below for $N$ up to $312$.  A clear intuitive interpretation of (\ref{overlap}) can be obtained by analogy to a simpler case.  Imagine that $\left| v_N \right>$ were, instead of a superconducting vortex state, a Hartree-Fock state of $N$ spinless fermions in a Hilbert space of total dimension $N_o ^N$.  In that case, $\left| v_N \right>$ would be a Slater determinant of $N$ orbitals, $\zeta_{i} ({\mathbf r})$ ($i=1,...,N$) drawn from a complete, orthonormal basis of $N_o$ orbitals, $\zeta_{i} ({\mathbf r})$ ($i = 1,...,N_o$).  It would follow from the definition of the Slater determinant that $\left< v_N(\mbox{\boldmath$\rho$}) \right. \left| v_N (\mbox{\boldmath$\rho$}^{\prime})\right> = \hbox{det}\; {\cal Z} \; ,$ for ${\cal Z}$ an $N \times N$ matrix of overlap integrals ${\cal Z}_{i,j} = \int dr \zeta _i^* ({\mathbf r} -\mbox{\boldmath$\rho$}) \zeta _j ({\mathbf r}-\mbox{\boldmath$\rho$}^\prime)   \; .$  Defining a related $N_o \times N_o$ matrix
\begin{equation*}
\tilde{\cal Z} = \left\{ \begin{array}{lr} \int dr \zeta _i^* ({\mathbf r} -\mbox{\boldmath$\rho$}) \zeta _j ({\mathbf r}-\mbox{\boldmath$\rho$}^\prime)& j = 1,\ldots,N \\ \delta _{i,j} & j = N+1,\ldots,N_o \end{array} \right.
\end{equation*}
we find that $\hbox{det} {\cal Z} = \hbox{det} \tilde{\cal Z}.$  Using the Green's functions \mbox{$G({\mathbf r},{\mathbf r^{\prime}},t=0^-) = \sum _{i \le N} \zeta _i ({\mathbf r}) \zeta ^* _i ({\mathbf r^{\prime}})$} and \mbox{$G({\mathbf r},{\mathbf r^{\prime}},t=0^+) = -\sum ^{N_o} _{i > N} \zeta _i ({\mathbf r}) \zeta ^* _i ({\mathbf r^{\prime}})$}, we construct the $N_o \times N_o$ matrix
\begin{equation*}
{\cal M}_{{\mathbf r},{\mathbf r^{\prime}}} = -G({\mathbf r}-\mbox{\boldmath$\rho$}^\prime,{\mathbf r^{\prime}}-\mbox{\boldmath$\rho$}^\prime,t=0^+) + G({\mathbf r}-\mbox{\boldmath$\rho$}^\prime,{\mathbf r}^{\prime}-\mbox{\boldmath$\rho$},t=0^-).
\end{equation*}
This matrix satisfies ${\cal U}^{\dagger} {\cal M} {\cal U} = \tilde{\cal Z}$ for the unitary matrix ${\cal U}_{{\mathbf r},i} \equiv \zeta _i ({\mathbf r} - \mbox{\boldmath$\rho$}^{\prime}) \; .$  Therefore, $\left< v_N(\mbox{\boldmath$\rho$}) \right. \left| v_N (\mbox{\boldmath$\rho$}^{\prime})\right> = \hbox{det}\; {\cal Z} = \hbox{det}\; \tilde{\cal Z} = \hbox{det}\; {\cal M}.$  The equation is reasonable: the $G({\mathbf r}-\mbox{\boldmath$\rho$}^{\prime},{\mathbf r^{\prime}}-\mbox{\boldmath$\rho$},t=0^-)$ part of ${\cal M}$ translates the $N$ occupied orbitals in $\left| v_N (\mbox{\boldmath$\rho$}^{\prime})\right> $ to $\mbox{\boldmath$\rho$}$ while the $-G({\mathbf r}-\mbox{\boldmath$\rho$}^{\prime},{\mathbf r^{\prime}}-\mbox{\boldmath$\rho$}^{\prime},t=0^+) = \delta({\mathbf r}-{\mathbf r^{\prime}}) - G({\mathbf r}-\mbox{\boldmath$\rho$}^{\prime},{\mathbf r^{\prime}}-\mbox{\boldmath$\rho$}^{\prime},t=0^-)$ part leaves the $N_o - N$ unoccupied orbitals untranslated.  Similar intuition applies to our superconducting case, although we have a magnetic field and we also integrate over phase $\chi$ to project $\left| v \right>$ on to the state $\left| v_N \right>$. 

The reasoning leading to equation (\ref{overlap}) also permits us to compute \cite{condmat} the matrix elements of the pinning potential 
\begin{eqnarray}
\lefteqn{\left< v_N (\mbox{\boldmath$\rho$}) \right| V \left| v_N (\mbox{\boldmath $\rho$}^\prime)\right> =  \left< v \right|T^\dagger(\mbox{\boldmath$\rho$}) V T(\mbox{\boldmath$\rho$}^\prime)P_N \left| v \right>}  \label{potential} \\
&= & \sum_{{\mathbf r}_o,s_o} V_{s_o}({\mathbf r}_o) \int \frac{d\chi}{2\pi}  e^{i(N_o/2-N)\chi}  \hbox{det} {\cal L} (\chi,{\mathbf r}_o,s_o) \notag
\end{eqnarray}
where 
\begin{widetext}
\begin{eqnarray*}
\lefteqn{{\cal L}(\chi,{\mathbf r}_o,s_o) _{({\mathbf r},s);({\mathbf r^{\prime}},s^{\prime})}  =  - G_{s,s^{\prime}}({\mathbf r} - \mbox{\boldmath$\rho^\prime$},{\mathbf r}^{\prime} - \mbox{\boldmath$\rho^\prime$},t=0^+)(1 - \delta_{({\mathbf r}^\prime,s^\prime);({\mathbf r}_o,s_o)}\delta_{s_o,1})}   \\
& & + G_{s,s^{\prime}}({\mathbf r} - \mbox{\boldmath$\rho^\prime$},{\mathbf r}^\prime - \mbox{\boldmath$\rho$},t=0^-)e^{is^{\prime}\left(\chi + (\pi B/\Phi_o) ({\rho _y}^\prime (x^\prime - {\rho _x}^\prime /2) - \rho _y (x^\prime - \rho _x/2))\right)} (1 - \delta_{({\mathbf r}^\prime,s^\prime);({\mathbf r}_o,s_o)}\delta_{s_o,-1})   \; .
\end{eqnarray*}
\end{widetext}

Equations (\ref{schroeqn}), (\ref{overlap}), and (\ref{potential}) provide the ingredients for a microscopic calculation of quantum vortex tunneling.  We stress that, although formulae (\ref{overlap}) and (\ref{potential}) are complicated in appearance, they are attractive and manageable from a computational standpoint.  No formula is needed for the remaining matrix element $\left< v_N(\mbox{\boldmath$\rho$}) \right| H_o \left| v_N(\mbox{\boldmath$\rho$}^{\prime}) \right>$ if we are willing to approximate that $H_o$ has $\left| v_N(\mbox{\boldmath$\rho$}) \right>$ as an eigenstate.  We then just get $\left< v_N(\mbox{\boldmath$\rho$}) \right| H_o \left| v_N(\mbox{\boldmath$\rho$}^\prime) \right> = E_v \left< v_N(\mbox{\boldmath$\rho$}) \right. \left| v_N(\mbox{\boldmath$\rho$}^{\prime}) \right> \;$  and the constant energy $E_v$ merely shifts the eigenvalue $E$ in (\ref{schroeqn}).

\begin{figure}[tbp]
\includegraphics[height=4cm,angle=0]{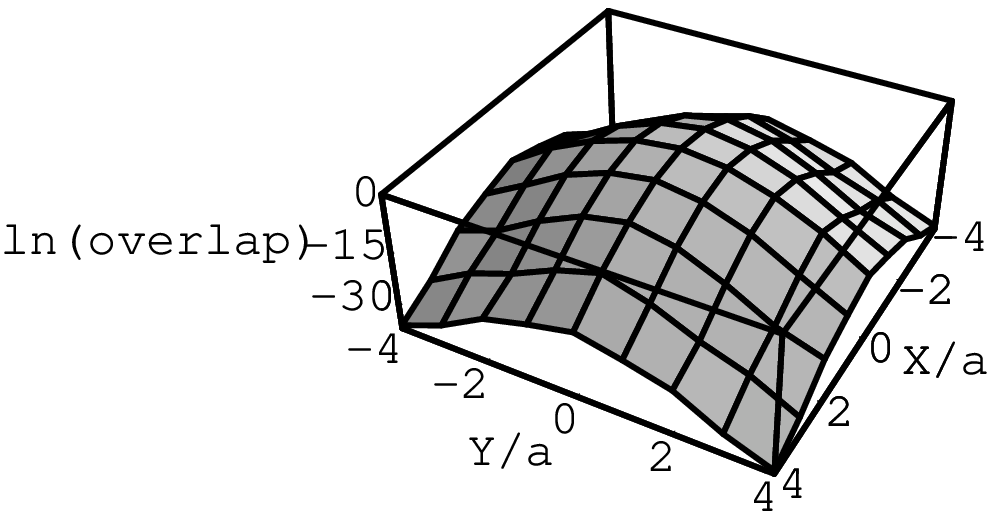} 
\caption{Magnitude of the overlap $|\left<v_N(\mbox{\boldmath$\rho$} = X\hat{x} + Y\hat{y}) \right. \left| v_N({\mathbf 0}) \right>|$ of two vortex states as a function of displacement $X\hat{x} + Y\hat{y}$ between the states.}
\label{overlapfigure}
\end{figure}

\begin{figure}[tbp]
\includegraphics[height=5.3cm,angle=0]{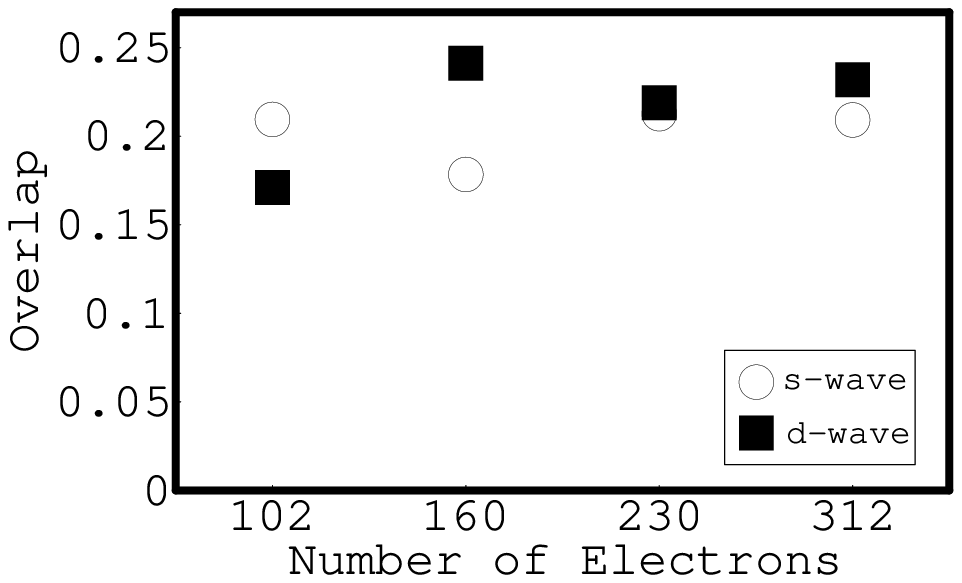}
\caption{Magnitude of nearest neighbor overlap $|\left<v_N(\mbox{\boldmath$\rho$} = a\hat{y}) \right. \left| v_N({\mathbf 0}) \right>|$ for calculations on lattices of increasing size, with $s$-wave and $d_{x^2-y^2}$ wave order parameters.}
\label{sdoverlapconverge}
\end{figure}

This formalism makes is possible to study, for instance, how
$d_{x^2-y^2}$-wave order parameter symmetry influences vortex quantum
tunneling in a high temperature superconductor.  We model a high
temperature superconductor with the Hamiltonian described in reference
\cite{Wang}.  We follow the methodology of \cite{Wang} to
obtain self-consistent quasiparticle amplitudes $(u,v)$ and then sum
these amplitudes to produce the self-consistent Nambu-Gorkov Green's
function appearing in (\ref{overlap}) and (\ref{potential}) -- this is
equivalent to solving the Nambu-Gorkov equations directly.
Calculations are made on a $14 \times 28$ real-space lattice ($N_o = 2 \times 14 \times 28 = 784$ single
particle states), where the number $N \approx 0.8 (N_o/2) = 312$ of
electrons in the superconductor is set near optimal doping.  Figure
(\ref{overlapfigure}) plots the magnitude of the calculated overlap
$|\left<v_N (\mbox{\boldmath$\rho$}= X\hat{x} +
Y\hat{y})\right. \left| v_N({\mathbf 0}) \right>|$ as a function of
displacement $X$ and $Y$ in units of the lattice constant $a$.
Results are shown for an $d_{x^2-y^2}$-wave superconductor; because it is a
logarithmic plot, the figure is not noticeably changed for
$s$-wave order parameter calculations. The results presented
in figure (\ref{overlapfigure}) do not depend sensitively upon the $14
\times 28$ lattice size used in the numerical calculation.  To
demonstrate this, figure (\ref{sdoverlapconverge}) displays the
nearest neighbor overlap $|\left<v_N (\mbox{\boldmath$\rho$} =
a\hat{y}) \right. \left| v_N ({\mathbf 0}\right>|$ for lattices of
increasing size.  Note that the magnitude of the overlap changes
little even when the number of electrons grows by more than a factor
of 3.  This finding confirms that interactions can lead to finite
vortex overlaps in infinite systems \cite{Niu}.  The result is in
contrast to the case of a system of $N$ non-interacting bosons, for
instance, in which the overlap decays exponentially with increasing
number of particles.

We model quantum vortex tunneling between two pinning sites by
inserting impurities into our self-consistent calculation of the
Green's function.  Of course, if our aim were to guide technological
efforts to raise critical currents, we would seek quantitative results
from the formalism by computing a realistic pinning potential $V$.  To
explore generic behavior, though, we take the pinning potential to be
a double well of well spacing $\mbox{\boldmath$\rho$}_o$, with depth
given by $1/7$ the Fermi energy, a reasonable magnitude.  We further
assume that the effect of this potential on the system's electrons is
to lower the energy of $\left| v_N ({\mathbf 0}) \right>$ and $\left|
v_N (\mbox{\boldmath$\rho$}_o) \right>$ below the energy of all other
states $\left| v_N (\mbox{\boldmath$\rho$}) \right>$.  Then the
Schr\"odinger equation (\ref{schroeqn}) is solved, in the two
dimensional Hilbert space spanned by these states.  If the
superconductor begins in a time-dependent state $\left| \Psi(t)
\right>$ satisfying $\left| \Psi(t=0) \right> = \left| v_N
(\mbox{\boldmath$\rho$}_o)\right>$, the probability that tunneling has
occured by time $t$ is $P(t) = 1 - \left| \left< v_N
(\mbox{\boldmath$\rho$}_o)\right. \left| \Psi(t) \right> \right| ^2 \;
.$ This probability is found to be periodic in our double well model,
and we therefore set the tunneling rate
$\Gamma(\mbox{\boldmath$\rho$}_o)$ equal to the reciprocal of the
period.  Although the model produces tunneling back to the initial
well, it is assumed, as in other macroscopic quantum tunneling models
\cite{Leggett}, that the environment intercedes to prevent this from
occuring.  Since the dissipative effects of the environment are not
explicity included in the calculation, we should obtain an upper bound
on the tunneling rate \cite{dissipation}.  One could include the
effects of dissipation in future work by recasting the dynamics of
(\ref{schroeqn}) in a functional integral as in \cite{Caldeira}.

As a measure of the directionality of the tunneling, we compute the
ratio $\Gamma(a\hat{x} + a\hat{y}) / \Gamma(a\hat{y})$, using formulae
(\ref{schroeqn}), (\ref{overlap}), and (\ref{potential}) to evaluate
$\Gamma(\mbox{\boldmath$\rho$}_o)$.  In the $s$-wave case for a $14
\times 28$ lattice, we obtain a ratio of $50 \%$.  In the
$d_{x^2-y^2}$-wave case for a $14 \times 28$ lattice, the ratio is $2
\%$.  In comparison to the $s$-wave case, the $d_{x^2-y^2}$-wave order
parameter symmetry strongly suppresses the relative tunneling rate
along the order parameter nodes.  When the high potential
region between the two wells is obstructing motion along the nodes of
the order parameter, it is more effective in confining the
superconducting state.

The effect could be probed by magnetization measurements of the kind reviewed in \cite{Yeshurun}.  A potential experimental arrangement involves the fabrication of two superconducting cylindrical tubes of square cross-section.  One cylinder has the nodes of the $d$-wave order parameter parallel to its walls; the other cylinder has the nodes angled at $45^\circ$ to its walls.  A vortex trapped in the hollow interior of a cylinder by field cooling will tunnel out at a rate influenced by the order parameter orientation.

\begin{acknowledgments}
The author thanks A. Auerbach, who initially suggested that the author study anisotropic dynamics of $d$-wave vortices, J. Jain, G. Koren, M. Wolraich and Y. Yeshurun.  The author is grateful for the hospitality of The Institute for Theoretical Physics at Technion during the early stages of this work and for funding provided by a Packard Foundation.
\end{acknowledgments}


\begin{thebibliography}{43}
\bibitem[{\citenamefont{{anistropy}}()}]{anisotropy}
\bibinfo{note}{This cannot be studied even by including an anisotropic effective mass tensor in a point-particle model: unless one introduces momentum dependence into the vortex mass, it must be isotropic in a $d$-wave superconductor since the superconductor has 90$^\circ$ rotational symmetry}.
\end{thebibliography}
\end{document}


Space constraints make it impossible to present detailed derivations of our results within the body of our manuscript.  This appendix supplies explicit mathematical arguments for the convenience of interested readers.

\section{Particle Hole Transformation}
To evaluate the overlap matrix element $\left< v_N (\mbox{\boldmath$\rho$}) \right. \left| v_N (\mbox{\boldmath$\rho$}^{\prime})\right>$, one inserts the appropriate translation operators $\left< v_N (\mbox{\boldmath$\rho$}) \right. \left| v_N (\mbox{\boldmath$\rho$}^{\prime})\right> =   \left< v \right|T^\dagger(\mbox{\boldmath$\rho$}) T(\mbox{\boldmath$\rho^\prime$})P_N \left| v \right>$. Note that there is no operator $P_N$ next to $T^\dagger(\mbox{\boldmath$\rho$})$; a second projection operator would not change the matrix element.

Now, the definition of the many-body translation operator is 
\begin{equation*}
T(\mbox{\boldmath$\rho$})= \Pi _{{\mathbf R},S} (C _{{\mathbf
R},S} C ^\dagger _{{\mathbf
R},S} + c^\dagger _{{\mathbf
R},S} C _{{\mathbf
R},S}) \Pi _{{\mathbf r},s} (c_{{\mathbf r}-\mbox{\boldmath$\rho$},s}c^\dagger_{{\mathbf r}-\mbox{\boldmath$\rho$},s} + e^{i (\pi B/\Phi_o) \rho _y (x - \rho _x /2)} C^{\dagger}_{{\mathbf r},s}c_{{\mathbf r}-\mbox{\boldmath$\rho$},s})
\end{equation*}
and it is convenient introduce a particle-hole transformation.  We write
\begin{eqnarray*}
\lefteqn{\Pi _{{\mathbf R}} (C _{{\mathbf R},-1} C ^\dagger _{{\mathbf R},-1} + c^\dagger _{{\mathbf R},-1} C _{{\mathbf R},-1}) \Pi _{{\mathbf r}} (c_{{\mathbf r}-\mbox{\boldmath$\rho$},-1}c^\dagger_{{\mathbf r}-\mbox{\boldmath$\rho$},-1} + e^{i (\pi B/\Phi_o) \rho _y (x -\rho _x /2)} C^{\dagger}_{{\mathbf r},-1}c_{{\mathbf r}-\mbox{\boldmath$\rho$},-1})} \\
& = & \Pi _{{\mathbf R}} C _{{\mathbf R},-1} \Pi _{{\mathbf R}} (C^{\dagger} _{{\mathbf R},-1} c  _{{\mathbf R},-1} - C^{\dagger} _{{\mathbf R},-1} C _{{\mathbf R},-1}) \Pi _{{\mathbf R}} c^\dagger _{{\mathbf R},-1} \\
& & \times \Pi _{{\mathbf r}} c_{{\mathbf r}-\mbox{\boldmath$\rho$},-1} \Pi _{{\mathbf r}} (c^\dagger _{{\mathbf r}-\mbox{\boldmath$\rho$},-1}C _{{\mathbf r},-1} e^{-i (\pi B/\Phi_o) \rho _y (x -\rho _x /2)} - c^{\dagger}_{{\mathbf r}-\mbox{\boldmath$\rho$},-1}c_{{\mathbf r}-\mbox{\boldmath$\rho$},-1}) \Pi _{{\mathbf r}} C^\dagger _{{\mathbf r},-1} e^{i (\pi B/\Phi_o) \rho _y (x -\rho _x /2)} \\
& = & \Pi _{{\mathbf R}} C _{{\mathbf R},-1} \Pi _{{\mathbf R}} ( C^{\dagger} _{{\mathbf R},-1} C _{{\mathbf R},-1}+ c  _{{\mathbf R},-1} C^{\dagger} _{{\mathbf R},-1}) \\
& & \times \Pi _{{\mathbf r}} (c^{\dagger}_{{\mathbf r}-\mbox{\boldmath$\rho$},-1}c_{{\mathbf r}-\mbox{\boldmath$\rho$},-1}+e^{-i (\pi B/\Phi_o) \rho _y (x -\rho _x /2)} C _{{\mathbf r},-1} c^\dagger _{{\mathbf r}-\mbox{\boldmath$\rho$},-1}) \Pi _{{\mathbf r}} C^\dagger _{{\mathbf r},-1} e^{i (\pi B/2\Phi_o) \rho _x\rho _y)} 
\end{eqnarray*}
The phase factor at the end arises since magnetic boundary conditions imply $\Pi _{{\mathbf r}} c _{{\mathbf r}-\mbox{\boldmath$\rho$},-1} e^{i (\pi B/\Phi_o) \rho _y (x -\rho _x /2)} = \Pi _{{\mathbf r}} c _{{\mathbf r},-1} e^{i (\pi B/2\Phi_o) \rho _y \rho _x }$.  Defining ``holding-space'' operators $D_{{\mathbf R},1} = C_{{\mathbf R},1}$ and $D_{{\mathbf R},-1} = C^{\dagger}_{{\mathbf R},-1}$, we therefore find 
\begin{eqnarray}
T(\mbox{\boldmath$\rho$}) & = & \Pi _{{\mathbf r}} e^{i(\pi B /2\Phi_o) \rho _x \rho _ y} (\Pi _{{\mathbf R}} D^\dagger _{{\mathbf R},-1})  \Pi _{{\mathbf R},S} (D _{{\mathbf R},S} D ^\dagger _{{\mathbf R},S} + d^\dagger _{{\mathbf R},S} D  _{{\mathbf R},S})\notag \\
& & \; \Pi _{{\mathbf r},s} (d_{{\mathbf r}-\mbox{\boldmath$\rho$},s}d^\dagger_{{\mathbf r}-\mbox{\boldmath$\rho$},s} + e^{is(\pi B/\Phi_o) \rho _y (x - \rho _x /2)}D^\dagger_{{\mathbf r},s}d_{{\mathbf r}-\mbox{\boldmath$\rho$},s}) (\Pi _{{\mathbf r}} D_{{\mathbf r},-1}) \; .\label{T}
\end{eqnarray}
The leading phase factor can be absorbed into the definition of the vortex state if desired, as $\left| v_N(\mbox{\boldmath$\rho$}) \right> \equiv \Pi _{{\mathbf r}} e^{-i(\pi B /\Phi_o) \rho _x \rho _ y/2} T(\mbox{\boldmath$\rho$}) \left| v_N \right>$, and can therefore be omitted from (\ref{T}).\

A nearly identical calculation yields the operator $T(\mbox{\boldmath$\rho$})P_N$, although an additional phase factor $e^{i(N_o/2)\chi}$ arises.  Knowing the translation operator, we can compute the overlap and translation matrix elements.

\section{Overlap Matrix Element}
The overlap matrix element is
\begin{eqnarray}
\lefteqn{\left< v_N (\mbox{\boldmath$\rho$}) \right. \left| v_N (\mbox{\boldmath$\rho$}^{\prime})\right>} \notag \\
& = & \; \int \frac{d \chi}{2\pi} e^{i(N_o/2 - N) \chi} \left< v \right| (\Pi _{{\mathbf r}} D^\dagger_{{\mathbf r},-1})  \Pi _{{\mathbf r},s} (d_{{\mathbf r}-\mbox{\boldmath$\rho$},s}d^\dagger_{{\mathbf r}-\mbox{\boldmath$\rho$},s} + e^{-is(\pi B/\Phi_o) \rho _y (x - \rho _x /2)} d^\dagger_{{\mathbf r}-\mbox{\boldmath$\rho$},s} D _{{\mathbf r},s}) \notag\\
& & \; \Pi _{{\mathbf R},S} (D _{{\mathbf R},S} D ^\dagger _{{\mathbf R},S} + D^\dagger _{{\mathbf R},S} d  _{{\mathbf R},S})   (\Pi _{{\mathbf R}} D _{{\mathbf R},-1}) \times  (\Pi _{{\mathbf R}^\prime} D^\dagger _{{\mathbf R}^\prime,-1})  \Pi _{{\mathbf R}^\prime,S^\prime} (D _{{\mathbf R}^\prime,S^\prime} D ^\dagger _{{\mathbf R}^\prime,S^\prime} + d^\dagger _{{\mathbf R}^\prime,S^\prime} D  _{{\mathbf R}^\prime,S^\prime}) \notag \\
& & \; \Pi _{{\mathbf r}^\prime,s^\prime} (d_{{\mathbf r}^\prime-\mbox{\boldmath$\rho^\prime$},s^\prime}d^\dagger_{{\mathbf r}^\prime-\mbox{\boldmath$\rho^\prime$},s^\prime} + e^{is^\prime\left(\chi + (\pi B/\Phi_o) \rho ^\prime _y (x ^\prime - \rho ^\prime_x /2)\right)} D^\dagger_{{\mathbf r}^\prime,s^\prime}d_{{\mathbf r}^\prime-\mbox{\boldmath$\rho^\prime$},s^\prime}) (\Pi _{{\mathbf r}^\prime} D_{{\mathbf r}^\prime,-1}) \left| v \right> \label{overlapstart} \\
& = & \int \frac{d\chi}{2\pi}  e^{i(N_o/2 -N)\chi}   \hbox{det} \left[ G_{s,s^{\prime}}({\mathbf r}-\mbox{\boldmath$\rho$}^\prime,{\mathbf r}^\prime - \mbox{\boldmath$\rho$},t=0^-)e^{is^{\prime}\left(\chi + (\pi B/\Phi_o) (\rho ^\prime _y (x^\prime - \rho ^\prime _x /2) - \rho _y (x^\prime - \rho _x /2)) \right)} \right. \notag \\ 
& & \left. - G_{s,s^{\prime}}({\mathbf r}-\mbox{\boldmath$\rho^\prime$},{\mathbf r}^{\prime}-\mbox{\boldmath$\rho^\prime$},t=0^+)  \right] \notag \\ 
& \equiv & \int \frac{d\chi}{2\pi}  e^{i(N_o/2 -N)\chi}  \hbox{det} \; {\cal M} (\chi). \label{overlapdone}
\end{eqnarray}

The equality of (\ref{overlapstart}) and (\ref{overlapdone}) is demonstrated as follows. We start by noticing that the middle factors $(\Pi _{{\mathbf R}} D _{{\mathbf R},-1})(\Pi _{{\mathbf R}^\prime} D^\dagger _{{\mathbf R}^\prime,-1})$ give unity when acting to the left or the right.  With these operators eliminated, we consider the new middle factors
\begin{eqnarray*}
\lefteqn{\Pi _{{\mathbf R},S}(D _{{\mathbf R},S} D ^\dagger _{{\mathbf R},S} + D^\dagger _{{\mathbf R},S} d  _{{\mathbf R},S}) \Pi _{{\mathbf R}^\prime,S^\prime} (D _{{\mathbf R}^\prime,S^\prime} D ^\dagger _{{\mathbf R}^\prime,S^\prime} + d^\dagger _{{\mathbf R}^\prime,S^\prime} D  _{{\mathbf R}^\prime,S^\prime}) = } \\
& & \;\;\;\;\;\;\; \Pi _{{\mathbf R},S}(D _{{\mathbf R},S} D ^\dagger _{{\mathbf R},S} + d^\dagger _{{\mathbf R},S} D  _{{\mathbf R},S} + D ^\dagger _{{\mathbf R},S} d  _{{\mathbf R},S} + D^\dagger _{{\mathbf R},S} D  _{{\mathbf R},S} d _{{\mathbf R},S} d ^\dagger _{{\mathbf R},S}) \rightarrow 1
\end{eqnarray*}
since the operator $d _{{\mathbf R},S}$ yields zero when it acts to the right, the operator $d^\dagger _{{\mathbf R},S}$ yields zero when it acts to the left, and the combination $d _{{\mathbf R},S}d ^\dagger _{{\mathbf R},S}$ yields unity.  This leaves
\begin{eqnarray*}
\left< v_N (\mbox{\boldmath$\rho$}) \right. \left| v_N (\mbox{\boldmath$\rho$}^{\prime})\right> & = & \int \frac{d \chi}{2\pi} e^{i(N_o/2 - N) \chi} \left< v \right| (\Pi _{{\mathbf r}} D^\dagger_{{\mathbf r},-1})  \Pi _{{\mathbf r},s} (d_{{\mathbf r}-\mbox{\boldmath$\rho$},s}d^\dagger_{{\mathbf r}-\mbox{\boldmath$\rho$},s} + e^{-is(\pi B/\Phi_o) \rho _y (x - \rho _x /2)} d^\dagger_{{\mathbf r}-\mbox{\boldmath$\rho$},s} D _{{\mathbf r},s}) \\
& & \; \Pi _{{\mathbf r}^\prime,s^\prime} (d_{{\mathbf r}^\prime-\mbox{\boldmath$\rho$}^\prime,s^\prime}d^\dagger_{{\mathbf r}^\prime-\mbox{\boldmath$\rho$}^\prime,s^\prime} +  e^{is^\prime\left(\chi + (\pi B/\Phi_o) \rho ^\prime _y (x ^\prime - \rho ^\prime_x /2)\right)} D^\dagger_{{\mathbf r}^\prime,s^\prime}d_{{\mathbf r}^\prime-\mbox{\boldmath$\rho^\prime$},s^\prime}) (\Pi _{{\mathbf r}^\prime} D_{{\mathbf r}^\prime,-1}) \left| v \right>  \\
& = & \int \frac{d \chi}{2\pi} e^{i(N_o/2 - N) \chi} \left< v \right| (\Pi _{{\mathbf r}} D^\dagger_{{\mathbf r},-1})  \Pi _{{\mathbf r},s} (1 + e^{-is(\pi B/\Phi_o) \rho _y (x - \rho _x /2)} d^\dagger_{{\mathbf r}-\mbox{\boldmath$\rho$},s} D _{{\mathbf r},s}) \\
& & \; \Pi _{{\mathbf r}^\prime,s^\prime} (d_{{\mathbf r}^\prime-\mbox{\boldmath$\rho$}^\prime,s^\prime}d^\dagger_{{\mathbf r}^\prime-\mbox{\boldmath$\rho$}^\prime,s^\prime} +  e^{is^\prime\left(\chi + (\pi B/\Phi_o) \rho ^\prime _y (x ^\prime - \rho ^\prime_x /2)\right)} D^\dagger_{{\mathbf r}^\prime,s^\prime}d_{{\mathbf r}^\prime-\mbox{\boldmath$\rho^\prime$},s^\prime}) (\Pi _{{\mathbf r}^\prime} D_{{\mathbf r}^\prime,-1}) \left| v \right>  
\end{eqnarray*}
Expanding the product out as a sum, one finds
\begin{eqnarray*}
& = & \int \frac{d\chi}{2\pi}  e^{i(N_o/2 -N)\chi} \sum _{n=0}^{N_o} \sum _{{\mathbf r}_1,s_1;...;{\mathbf r}_n,s_n}  \left< v \right| \left(e^{-is_1(\pi B/\Phi_o) \rho _y (x_1 - \rho _x /2)} d^{\dagger} _{{\mathbf r}_1 - \mbox{\boldmath$\rho$},s_1} \dots \right. \\
& & \left. \dots  e^{-is_n (\pi B/\Phi_o) \rho _y (x_n - \rho _x /2)} d^{\dagger} _{{\mathbf r}_n - \mbox{\boldmath$\rho$},s_n} \right)  \left(e^{is_1\left(\chi + (\pi B/\Phi_o) \rho ^\prime _y (x _1 - \rho ^\prime _x /2) \right)} d _{{\mathbf r}_1- \mbox{\boldmath$\rho^\prime$},s_1} \dots \right. \\
& & \left. e^{is_n\left(\chi + (\pi B/\Phi_o) \rho ^\prime _y (x _n - \rho ^\prime _x /2) \right)} d _{{\mathbf r}_n- \mbox{\boldmath$\rho^\prime$},s_n} \right) \left( d_{{\mathbf r}_{n+1}- \mbox{\boldmath$\rho^\prime$},s_{n+1}} d^\dagger _{{\mathbf r}_{n+1}- \mbox{\boldmath$\rho^\prime$},s_{n+1}} \dots d_{{\mathbf r}_{N_o}- \mbox{\boldmath$\rho^\prime$},s_{N_o}} d^\dagger _{{\mathbf r}_{N_o}- \mbox{\boldmath$\rho^\prime$},s_{N_o}} \right) \left| v \right>
\end{eqnarray*}
Invoking Wick's theorem, this leaves us with a sum over permutations $P$
\begin{eqnarray*}
& = & \int \frac{d\chi}{2\pi}  e^{i(N_o/2 -N)\chi}  \sum _P (-1)^P  \sum _{n=0}^{N_o} \sum _{{\mathbf r}_1,s_1;...;{\mathbf r}_n,s_n} \left( e^{is_1\left(\chi + (\pi B/\Phi_o)(\rho ^\prime _y (x _1 - \rho ^\prime _x /2) - \rho _y (x_1 - \rho _x /2)) \right) } \left< v \right| d^{\dagger} _{{\mathbf r}_1 - \mbox{\boldmath$\rho$},s_1} d _{{\mathbf r}_{P_1} - \mbox{\boldmath$\rho^\prime$},s_{P_1}} \left|v \right> ... \right.\\
& &\; \left. e^{is_n\left(\chi + (\pi B/\Phi_o)(\rho ^\prime _y (x _n - \rho ^\prime _x /2) - \rho _y (x_n - \rho _x /2)) \right) } \left< v \right| d^{\dagger} _{{\mathbf r}_n - \mbox{\boldmath$\rho$},s_n} d _{{\mathbf r}_{P_n} - \mbox{\boldmath$\rho^\prime$},s_{P_n}} \left|v \right> \right) \notag \\
& &\; \left< v \right| d  _{{\mathbf r}_{P_{n+1}- \mbox{\boldmath$\rho^\prime$}},s_{P_{n+1}}} d^\dagger _{{\mathbf r}_{n+1}- \mbox{\boldmath$\rho^\prime$},s_{n+1}} \left|v \right> ...  \left< v \right| d _{{\mathbf r}_{P_{N_o}}- \mbox{\boldmath$\rho^\prime$},s_{P_{N_o}}} d^\dagger _{{\mathbf r}_{N_o}- \mbox{\boldmath$\rho^\prime$},s_{N_o}} \left|v \right> \\
& = & \int \frac{d\chi}{2\pi}  e^{i(N_o/2 -N)\chi}   \sum _P (-1)^P \Pi_{i} \left(e^{is_i\left(\chi + (\pi B/\Phi_o) (\rho ^\prime _y (x _n - \rho ^\prime _x /2) - \rho _y (x_n - \rho _x /2))\right)} \left< v \right| d^{\dagger} _{{\mathbf r}_i - \mbox{\boldmath$\rho$},s_i} d _{{\mathbf r}_{P_i}- \mbox{\boldmath$\rho^\prime$},s_{P_i}} \left|v \right> \right.\\
&&\left.+ \left< v \right| d _{{\mathbf r}_{P_i}- \mbox{\boldmath$\rho^\prime$},s_{P_i}} d^\dagger _{{\mathbf r}_i- \mbox{\boldmath$\rho^\prime$},s_i} \left|v \right> \right) \\
& = & \int \frac{d\chi}{2\pi}  e^{i(N_o/2 -N)\chi}   \hbox{det} \; \left[ G_{s,s^{\prime}}({\mathbf r}- \mbox{\boldmath$\rho^\prime$},{\mathbf r}^\prime - \mbox{\boldmath$\rho$},t=0^-)e^{is^{\prime}\left(\chi +  (\pi B/\Phi_o) (\rho ^\prime _y (x ^\prime - \rho ^\prime _x /2) - \rho _y (x^\prime - \rho _x /2)) \right)} \right. \\
&& \left. - G_{s,s^{\prime}}({\mathbf r}- \mbox{\boldmath$\rho^\prime$},{\mathbf r}^{\prime}- \mbox{\boldmath$\rho^\prime$},t=0^+)\right] \\ 
& = & \int \frac{d\chi}{2\pi}  e^{i(N_o/2 -N)\chi}  \hbox{det} \; {\cal M} (\chi)
\end{eqnarray*}
The symbol $(-1)^P$ takes the value $+1$ for even permutations and $-1$ for odd permutations.  In the second to last equality, we recognize that a sum over permutations can be written as a determinant.

\section{Matrix Element of the Potential}

The analysis that enable us to compute $\left< v_N (\mbox{\boldmath$\rho$}) \right. \left| v_N (\mbox{\boldmath$\rho$}^{\prime})\right>$ can be extended to obtain matrix elements of the pinning potential  $\left< v_N (\mbox{\boldmath$\rho$}) \right| V  \left| v_N (\mbox{\boldmath$\rho$}^{\prime})\right>  = \; \left< v \right| T^{\dagger} (\mbox{\boldmath$\rho$}) V T(\mbox{\boldmath$\rho$}^{\prime}) P_N \left| v \right>$.

First, the potential is expressed in terms of the $d$ operators
\begin{eqnarray}
V & = & \sum_{{\mathbf r}_o,s_o} V_{s_o}({\mathbf r}_o) c^{\dagger}_{{\mathbf r}_o,s_o} c_{{\mathbf r}_o,s_o} =  \sum_{{\mathbf r}_o} V_{-1}({\mathbf r}_o) + \sum_{{\mathbf r}_o,s_o} s_o V_{s_o}({\mathbf r}_o) d^{\dagger}_{{\mathbf r}_o,s_o} d_{{\mathbf r}_o,s_o} \label{pinningpotential}
\end{eqnarray}
Given the form (\ref{pinningpotential}), it is necessary to compute $\left< v_N (\mbox{\boldmath$\rho$}) \right|d^{\dagger}_{{\mathbf r}_o,s_o} d_{{\mathbf r}_o,s_o} \left| v_N (\mbox{\boldmath$\rho$}^{\prime})\right>$.  The result is
\begin{eqnarray*}
\lefteqn{\left< v_N (\mbox{\boldmath$\rho$}) \right| d^{\dagger}_{{\mathbf r}_o,s_o} d_{{\mathbf r}_o,s_o} \left| v_N (\mbox{\boldmath$\rho$}^{\prime})\right>}\\
& = & \; \int \frac{d \chi}{2\pi} e^{i(N_o/2 - N) \chi} \left< v \right| (\Pi _{{\mathbf r}} D^\dagger_{{\mathbf r},-1})  \Pi _{{\mathbf r},s} (d_{{\mathbf r}-\mbox{\boldmath$\rho$},s}d^\dagger_{{\mathbf r}-\mbox{\boldmath$\rho$},s} + e^{-is(\pi B/\Phi_o) \rho _y (x - \rho _x /2)} d^\dagger_{{\mathbf r}-\mbox{\boldmath$\rho$},s} D _{{\mathbf r},s}) \notag\\
& & \; \Pi _{{\mathbf R},S} (D _{{\mathbf R},S} D ^\dagger _{{\mathbf R},S} + D^\dagger _{{\mathbf R},S} d  _{{\mathbf R},S})   (\Pi _{{\mathbf R}} D _{{\mathbf R},-1})d^{\dagger}_{{\mathbf r}_o,s_o} d_{{\mathbf r}_o,s_o} (\Pi _{{\mathbf R}^\prime} D^\dagger _{{\mathbf R}^\prime,-1})  \Pi _{{\mathbf R}^\prime,S^\prime} (D _{{\mathbf R}^\prime,S^\prime} D ^\dagger _{{\mathbf R}^\prime,S^\prime} + d^\dagger _{{\mathbf R}^\prime,S^\prime} D  _{{\mathbf R}^\prime,S^\prime}) \notag \\
& & \; \Pi _{{\mathbf r}^\prime,s^\prime} (d_{{\mathbf r}^\prime-\mbox{\boldmath$\rho^\prime$},s^\prime}d^\dagger_{{\mathbf r}^\prime-\mbox{\boldmath$\rho^\prime$},s^\prime} + e^{is^\prime\left(\chi + (\pi B/\Phi_o) \rho ^\prime _y (x ^\prime - \rho ^\prime_x /2)\right)} D^\dagger_{{\mathbf r}^\prime,s^\prime}d_{{\mathbf r}^\prime-\mbox{\boldmath$\rho^\prime$},s^\prime}) (\Pi _{{\mathbf r}^\prime} D_{{\mathbf r}^\prime,-1}) \left| v \right> \\
& = & \int \frac{d \chi}{2\pi} e^{i(N_o/2 - N) \chi} \left< v \right| (\Pi _{{\mathbf r}} D^\dagger_{{\mathbf r},-1})  \Pi _{{\mathbf r},s} (1 + e^{-is(\pi B/\Phi_o) \rho _y (x - \rho _x /2)} d^\dagger_{{\mathbf r}-\mbox{\boldmath$\rho$},s} D _{{\mathbf r},s}) D^\dagger _{{\mathbf r}_o,s_o}D _{{\mathbf r}_o,s_o}\\
& & \; \Pi _{{\mathbf r}^\prime,s^\prime} (d_{{\mathbf r}^\prime-\mbox{\boldmath$\rho$}^\prime,s^\prime}d^\dagger_{{\mathbf r}^\prime-\mbox{\boldmath$\rho$}^\prime,s^\prime} +  e^{is^\prime\left(\chi + (\pi B/\Phi_o) \rho ^\prime _y (x ^\prime - \rho ^\prime_x /2)\right)} D^\dagger_{{\mathbf r}^\prime,s^\prime}d_{{\mathbf r}^\prime-\mbox{\boldmath$\rho^\prime$},s^\prime}) (\Pi _{{\mathbf r}^\prime} D_{{\mathbf r}^\prime,-1}) \left| v \right>  
\end{eqnarray*}
From this calculation, we find that
\begin{eqnarray*}
\lefteqn{\left< v_N (\mbox{\boldmath$\rho$}) \right| V \left| v_N (\mbox{\boldmath$\rho$}^{\prime})\right>} \\
& = & \sum_{{\mathbf r}_o,s_o} \int \frac{d\chi}{2\pi}  e^{i(N_o/2 -N)\chi}   \sum _P (-1)^P \\
& &\; V_{-1}({\mathbf r}_o) \Pi_{{\mathbf r}_i,s_i} \left(e^{is_i\left(\chi + (\pi B/\Phi_o) (\rho ^\prime _y (x _n - \rho ^\prime _x /2) - \rho _y (x_n - \rho _x /2))\right)} \left< v \right| d^{\dagger} _{{\mathbf r}_i - \mbox{\boldmath$\rho$},s_i} d _{{\mathbf r}_{P_i}- \mbox{\boldmath$\rho^\prime$},s_{P_i}} \left|v \right> + \left< v \right| d _{{\mathbf r}_{P_i}- \mbox{\boldmath$\rho^\prime$},s_{P_i}} d^\dagger _{{\mathbf r}_i- \mbox{\boldmath$\rho^\prime$},s_i} \left|v \right> \right)  \\
& &\; - V_{-1}({\mathbf r}_o) \left( e^{-i\left(\chi + (\pi B/\Phi_o) (\rho ^\prime _y (x _o - \rho ^\prime _x /2) - \rho _y (x_o - \rho _x /2))\right)} \left< v \right| d^{\dagger} _{{\mathbf r}_o - \mbox{\boldmath$\rho$},-1} d _{{\mathbf r}_{P_o}- \mbox{\boldmath$\rho^\prime$},s_{P_o}} \left|v \right> \right) \\
& &\; \Pi_{{\mathbf r}_i,s_i \not= {\mathbf r}_o,-1} \left(e^{is_i\left(\chi + (\pi B/\Phi_o) (\rho ^\prime _y (x _n - \rho ^\prime _x /2) - \rho _y (x_n - \rho _x /2))\right)} \left< v \right| d^{\dagger} _{{\mathbf r}_i - \mbox{\boldmath$\rho$},s_i} d _{{\mathbf r}_{P_i}- \mbox{\boldmath$\rho^\prime$},s_{P_i}} \left|v \right> + \left< v \right| d _{{\mathbf r}_{P_i}- \mbox{\boldmath$\rho^\prime$},s_{P_i}} d^\dagger _{{\mathbf r}_i- \mbox{\boldmath$\rho^\prime$},s_i} \left|v \right> \right) \\
& &\; +  V_{1}({\mathbf r}_o)\left( e^{i\left(\chi + (\pi B/\Phi_o) (\rho ^\prime _y (x _o - \rho ^\prime _x /2) - \rho _y (x_o - \rho _x /2))\right)} \left< v \right| d^{\dagger} _{{\mathbf r}_o - \mbox{\boldmath$\rho$},1} d _{{\mathbf r}_{P_o}- \mbox{\boldmath$\rho^\prime$},s_{P_o}} \left|v \right> \right) \\
& &\; \Pi_{{\mathbf r}_i,s_i \not= {\mathbf r}_o,1} \left(e^{is_i\left(\chi + (\pi B/\Phi_o) (\rho ^\prime _y (x _n - \rho ^\prime _x /2) - \rho _y (x_n - \rho _x /2))\right)} \left< v \right| d^{\dagger} _{{\mathbf r}_i - \mbox{\boldmath$\rho$},s_i} d _{{\mathbf r}_{P_i}- \mbox{\boldmath$\rho^\prime$},s_{P_i}} \left|v \right> + \left< v \right| d _{{\mathbf r}_{P_i}- \mbox{\boldmath$\rho^\prime$},s_{P_i}} d^\dagger _{{\mathbf r}_i- \mbox{\boldmath$\rho^\prime$},s_i} \left|v \right> \right)  \\
& = & \sum_{{\mathbf r}_o,s_o} \int \frac{d\chi}{2\pi}  e^{i(N_o/2 -N)\chi}   \sum _P (-1)^P \\
& &\; V_{-1}({\mathbf r}_o) \left( \left< v \right| d _{{\mathbf r}_{P_o}- \mbox{\boldmath$\rho^\prime$},s_{P_o}} d^\dagger _{{\mathbf r}_o- \mbox{\boldmath$\rho^\prime$},s_o} \left|v \right>\right) \\
& &\; \Pi_{{\mathbf r}_i,s_i \not= {\mathbf r}_o,1} \left(e^{is_i\left(\chi + (\pi B/\Phi_o) (\rho ^\prime _y (x _n - \rho ^\prime _x /2) - \rho _y (x_n - \rho _x /2))\right)} \left< v \right| d^{\dagger} _{{\mathbf r}_i - \mbox{\boldmath$\rho$},s_i} d _{{\mathbf r}_{P_i}- \mbox{\boldmath$\rho^\prime$},s_{P_i}} \left|v \right> + \left< v \right| d _{{\mathbf r}_{P_i}- \mbox{\boldmath$\rho^\prime$},s_{P_i}} d^\dagger _{{\mathbf r}_i- \mbox{\boldmath$\rho^\prime$},s_i} \left|v \right> \right)  \\
& &\; +  V_{1}({\mathbf r}_o)\left(e^{i\left(\chi + (\pi B/\Phi_o) (\rho ^\prime _y (x _o - \rho ^\prime _x /2) - \rho _y (x_o - \rho _x /2))\right)} \left< v \right| d^{\dagger} _{{\mathbf r}_o - \mbox{\boldmath$\rho$},1} d _{{\mathbf r}_{P_o}- \mbox{\boldmath$\rho^\prime$},s_{P_o}} \left|v \right>\right) \\
& &\; \Pi_{{\mathbf r}_i,s_i \not= {\mathbf r}_o,1} \left(e^{is_i\left(\chi + (\pi B/\Phi_o) (\rho ^\prime _y (x _n - \rho ^\prime _x /2) - \rho _y (x_n - \rho _x /2))\right)} \left< v \right| d^{\dagger} _{{\mathbf r}_i - \mbox{\boldmath$\rho$},s_i} d _{{\mathbf r}_{P_i}- \mbox{\boldmath$\rho^\prime$},s_{P_i}} \left|v \right> + \left< v \right| d _{{\mathbf r}_{P_i}- \mbox{\boldmath$\rho^\prime$},s_{P_i}} d^\dagger _{{\mathbf r}_i- \mbox{\boldmath$\rho^\prime$},s_i} \left|v \right> \right)  \\
& = & \sum_{{\mathbf r}_o,s_o} V_{s_o}({\mathbf r}_o ) \int \frac{d\chi}{2\pi}  e^{i(N_o/2-N)\chi}  \hbox{det} {\cal L} (\chi,{\mathbf r}_o,s_o)
\end{eqnarray*}
where 
\begin{eqnarray*}
{\cal L}(\chi,{\mathbf r}_o,s_o) _{({\mathbf r},s);({\mathbf r^{\prime}},s^{\prime})}  & = &    G_{s,s^{\prime}}({\mathbf r} - \mbox{\boldmath$\rho^\prime$},{\mathbf r}^\prime - \mbox{\boldmath$\rho$},t=0^-)e^{is^{\prime}\left(\chi + (\pi B/\Phi_o) ({\rho _y}^\prime (x^\prime - {\rho _x}^\prime /2) - \rho _y (x^\prime - \rho _x/2))\right)} (1 - \delta_{({\mathbf r}^\prime,s^\prime);({\mathbf r}_o,s_o)}\delta_{s_o,-1})\\
& & -  G_{s,s^{\prime}}({\mathbf r} - \mbox{\boldmath$\rho^\prime$},{\mathbf r}^{\prime} - \mbox{\boldmath$\rho^\prime$},t=0^+) (1 - \delta_{({\mathbf r}^\prime,s^\prime);({\mathbf r}_o,s_o)}\delta_{s_o,1})\; .
\end{eqnarray*}
